\documentclass[journal]{IEEEtran}

\usepackage{booktabs}
\usepackage{amsmath}
\usepackage{amsthm}
\usepackage{multirow}
\usepackage{color,cite}
\usepackage{subfigure}
\usepackage{mathrsfs}
\usepackage{amssymb}
\usepackage{bm}
\usepackage{graphicx}
\usepackage{amsfonts}
\usepackage{algorithm}
\usepackage{algorithmic}
\usepackage{epstopdf}

\usepackage[top=1.92cm, bottom=2.5cm, left=1.5cm, right=1.5cm]{geometry}

\begin{document}

\bibliographystyle{IEEEtran}

\vspace{-10mm}

\title{\huge{\huge{TypeII-CsiNet: CSI Feedback with TypeII Codebook}}
\vspace{-2mm}}

\author{
\IEEEauthorblockN{Yiliang Sang$^{\rm 1}$, Ke Ma$^{\rm 1}$, Yang Ming$^{\rm 2}$, Jin Lian$^{\rm 2}$, and Zhaocheng Wang$^{\rm 1, \rm 3}$, \emph{Fellow, IEEE}} \\%
\IEEEauthorblockA{$^{\rm 1}$Beijing National Research Center for Information Science and Technology (BNRist), \\
 Department of Electronic Engineering, Tsinghua University, Beijing 100084, China\\
 $^{\rm 2}$Huawei Technologies Company, Ltd., Beijing 100095, China \\
 $^{\rm 3}$Shenzhen International Graduate School, Tsinghua University, Shenzhen 518055, China\\
\{sangyl23, ma-k19, mingy20\}@mails.tsinghua.edu.cn, lianjin7@huawei.com, zcwang@tsinghua.edu.cn}
\vspace{-9mm}}

\maketitle
\thispagestyle{empty}

\begin{abstract}
The latest TypeII codebook selects partial strongest angular-delay ports for the feedback of downlink channel state information (CSI), whereas its performance is limited due to the deficiency of utilizing the correlations among the port coefficients. To tackle this issue, we propose a tailored autoencoder named TypeII-CsiNet to effectively integrate the TypeII codebook with deep learning, wherein three novel designs are developed for sufficiently boosting the sum rate performance. Firstly, a dedicated pre-processing module is designed to sort the selected ports for reserving the correlations of their corresponding coefficients. Secondly, a position-filling layer is developed in the decoder to fill the feedback coefficients into their ports in the recovered CSI matrix, so that the corresponding angular-delay-domain structure is adequately leveraged to enhance the reconstruction accuracy. Thirdly, a two-stage loss function is proposed to improve the sum rate performance while avoiding the trapping in local optimums during model training. Simulation results verify that our proposed TypeII-CsiNet outperforms the TypeII codebook and existing deep learning benchmarks. 
\end{abstract}

\IEEEpeerreviewmaketitle

\vspace*{-5mm}

\section{Introduction}\label{Sec:INTRO} 

Multiple-input multiple-output (MIMO) is widely regarded as one of the key technologies for supporting multi-user (MU) scenarios in the next-generation wireless communication systems. To sufficiently leverage the benefits of MIMO in frequency division duplex (FDD) systems, each user equipment (UE) is required to feed back its estimated downlink (DL) channel state information (CSI) to the base station (BS) via uplink (UL) channels. However, the full CSI matrix is so huge that the corresponding feedback overhead becomes unacceptable due to the substantial antennas.

To deal with this problem, the TypeII codebook was standardized by the 3rd generation partnership project (3GPP) for supporting the low-overhead feedback of CSI eigenvectors \cite{1}, whereas the full CSI matrix needs to be measured at UE side. To further reduce the overhead of DL CSI measurement, the partial reciprocity between UL and DL channels was exploited in the latest TypeII codebook \cite{2}, where the BS selects partial strongest angular-delay ports of the available UL CSI matrix for designing DL pilots. Accordingly, the UE only estimates and feeds back the coefficients of the selected ports to reduce the overhead of both DL pilots and UL feedback. However, the latest TypeII codebook suffers from two limitations. Firstly, before feedback, the quantization is performed for each port coefficient separately to obtain the bitstream, whereas the correlations among the port coefficients are neglected. Secondly, only the coefficients of the selected ports are reconstructed at BS side, while the coefficients of the unselected ports that can be recovered based on the correlations are ignored, so that the quality of CSI reconstruction is restricted.

Fortunately, deep learning can adaptively utilize the correlations among the port coefficients for efficient compression and reconstruction, hence CSI feedback with deep learning has been widely studied in recent years \cite{3,4,5,6,7,8,9,10,11,12}. Specifically, the existing works mainly focused on the feedback of the full CSI matrix by the autoencoder (AE) \cite{3,4,5,6,7,8}. By contrast, a few works followed the 3GPP standard and paid attention to the feedback of CSI eigenvectors with deep learning \cite{9,10,11,12}. The work \cite{9} proposed an AE named EVCsiNet, wherein the eigenvector decomposition is performed on the full CSI matrix before compression at UE side. Besides, a bidirectional long short-term memory network was introduced for further leveraging the frequency correlations \cite{10}. Moreover, MixerNet was proposed in \cite{11} to sufficiently utilize both the spatial and frequency correlations, while the dedicated design for the structure of dual-polarized antennas was investigated in \cite{12}. However, the full CSI matrix is required to be estimated at UE side in the above studies, which leads to the enormous overhead of DL pilots.

Thanks to the partial reciprocity exploited in the latest TypeII codebook, integrating the TypeII codebook with deep learning provides the feasibility to utilize the correlations among the port coefficients for enhancing the feedback performance, while guaranteeing the low overhead of DL pilots. However, in the latest TypeII codebook-based CSI feedback, the selected angular-delay ports in the CSI matrix are unknown to the UE, which destroys the stability of the correlations in the feedback coefficient vector. This characteristic could hamper the performance of the AE, since deep learning is not good at extracting the features from non-sparse data \cite{13,14,15}. 

In this paper, a tailored AE structure called TypeII-CsiNet is proposed to effectively integrate the latest TypeII codebook with deep learning, wherein three novel designs are developed for sufficiently boosting the sum rate performance in MU-MIMO scenarios. Firstly, after port selection at BS side, a dedicated sorting module is designed to shift the selected ports to the stable positions for reserving the correlations in the feedback coefficient vector. Secondly, a position-filling layer is deployed in the decoder to fill the feedback coefficients into their corresponding ports in the recovered CSI matrix. In this way, the angular-delay-domain structure of the recovered CSI matrix is fully leveraged, hence the decoder can efficiently exploit the correlations to improve the reconstruction accuracy. Thirdly, we propose a two-stage loss function for training the proposed TypeII-CsiNet. Concretely, in the first stage, the mean square error (MSE) loss function is applied for the coarse reconstruction. In the second stage, we propose a mixed loss function to combine the advantages of the supervised MSE and unsupervised rate loss functions for simultaneously avoiding the trapping in local optimums and improving the sum rate performance. Simulation results verify that our proposed TypeII-CsiNet outperforms the TypeII codebook and existing deep learning benchmarks in terms of the sum rate performance.


\section{System Model}\label{Sec:SYS} 


\subsection{DL Transmission Model}\label{S2.1} 

Consider a single-cell FDD MU-MIMO system where a BS serves $K$ single-antenna UEs, while our proposed scheme can be straightforwardly extended to the scenario with multiple UE antennas. The BS is equipped with dual-polarized antennas in the uniform planar array (UPA), where the numbers of antennas in the horizontal and vertical directions are $N_\text{h}$ and $N_\text{v}$. Thus, the total number of antennas at the BS is $N_\text{t} = 2 N_\text{h} N_\text{v}$. Following the 3GPP standard \cite{2}, we denote $M$ as the number of subbands, where each subband contains $N_\text{R}$ resource blocks (RBs) as the basic feedback granularity, and each RB consists of $12$ subcarriers in orthogonal frequency division multiplexing systems. For the $k$-th UE, the DL received signal on the $m$-th subband $y_{m, k}$ can be written as
\begin{equation}\label{Eq:channel_model} 
y_{m, k} = \sqrt{P} {\bm{h}^\text{H}_{m, k}}  \bm{v}_{m, k} x_{m, k} + \sum_{j \neq k} \sqrt{P} {\bm{h}^\text{H}_{m, k}} \bm{v}_{m, j} x_{m, j} + z_{m, k}.
\end{equation}
In (\ref{Eq:channel_model}), the first term denotes the signal of the $k$-th UE on the $m$-th subband, wherein $\bm{h}_{m, k} \in \mathbb{C}^{N_\text{t} \times 1}$ is the DL channel vector, $\bm{v}_{m, k} \in \mathbb{C}^{N_\text{t} \times 1}$ represents the precoding vector satisfying $\big\|\bm{v}_{m, k}\big\|^2 = 1$, $x_{m, k}$ is the transmitted symbol with $|x_{m, k}|=1$, and $P$ is the transmit power. The second term denotes the corresponding interference from other UEs. The third term $z_{m, k}$ represents the additional white Gaussian noise (AWGN) with the noise power $\sigma^2$, i.e., $z_{m, k} \sim \mathcal{CN} \big(0,\sigma^2 \big)$. Generally, the sum rate $R_\text{SUM}$ is utilized to evaluate the performance of the MU-MIMO system, which is written as
\begin{equation}\label{Eq:SR} 
R_\text{SUM}= \frac{1}{M} \sum_{m=1}^{M} \sum_{k=1}^{K} \log_2(1 + \frac{ P \big|{\bm{h}^\text{H}_{m, k}} \bm{v}_{m, k}\big| ^2} { \sum_{j \neq k} P \big|{\bm{h}^\text{H}_{m, k}} \bm{v}_{m, j}\big| ^2 + \sigma^2 } ).
\end{equation}
For convenience, the DL CSI matrix of the $k$-th UE on all subbands is represented as $\bm{H}_{k, \text{DL}}\! =\! \big[\bm{h}_{1,k} ~ \bm{h}_{2,k} \cdots \bm{h}_{M, k}\big] \in \mathbb{C}^{N_\text{t} \times M}$. Since the feedback overhead of $\bm{H}_{k, \text{DL}}$ containing $N_\text{t} M$ elements is excessive, $\bm{H}_{k, \text{DL}}$ needs to be compressed.

\vspace*{-2mm}

\subsection{TypeII Codebook}\label{S2.2} 

\begin{figure}[tp!]
\vspace*{-1mm}
\begin{center}
\includegraphics[width=.45\textwidth]{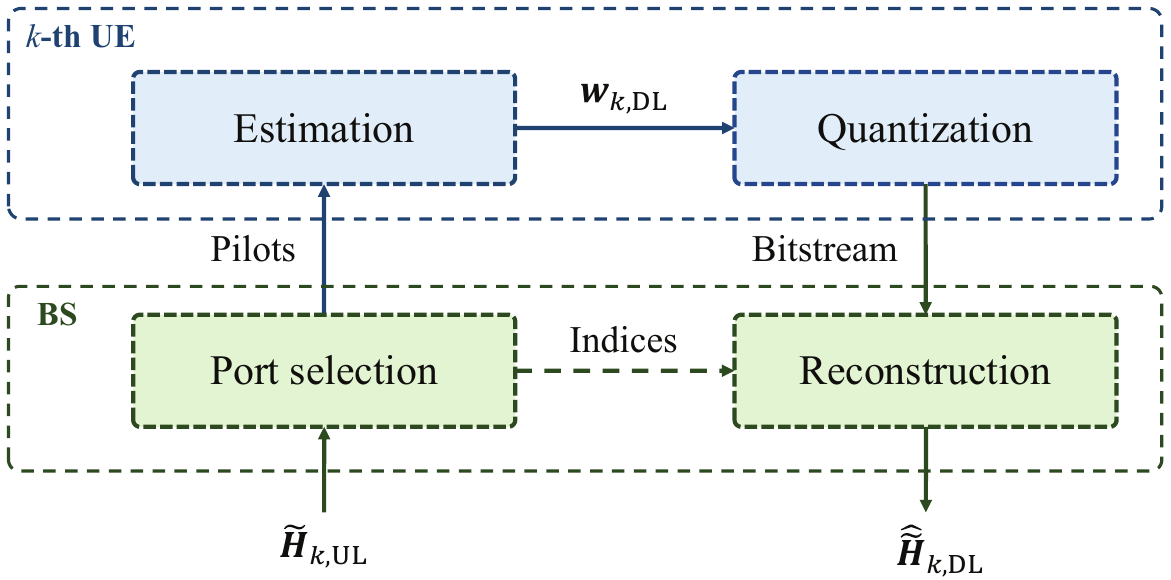}
\end{center}
\vspace{-4mm}
\caption{Framework of the latest TypeII codebook.}
\label{Fig:feedback_model} 
\vspace*{-5mm}
\end{figure}

The framework of the latest TypeII codebook \cite{2} is depicted in Fig.~\ref{Fig:feedback_model}, which is divided into four steps, port selection, estimation, quantization, and reconstruction. Since we focus on the feedback design, the estimation step is assumed to be perfect \cite{6,9}, which is omitted in the following. 

\textbf{Port selection}: The UL CSI matrix of the $k$-th UE $\bm{H}_{k, \text{UL}} \in \mathbb{C}^{N_\text{t} \times M}$ can be estimated at the BS and then transformed from the spatial-frequency domain to the angular-delay domain as follows
\begin{equation}\label{Eq:UL_DFT} 
\widetilde{\bm{H}}_{k, \text{UL}} = \bm{F}_\text{A}^{\text{H}} \bm{H}_{k, \text{UL}} \bm{F}_\text{D},
\end{equation}
where $\widetilde{\bm{H}}_{k, \text{UL}} \in \mathbb{C}^{N_\text{t} \times M}$ represents the corresponding UL CSI matrix in the angular-delay domain. $\bm{F}_\text{A} = \text{diag}(\bm{B}, \bm{B}) \in \mathbb{C}^{N_{\text{t}} \times N_{\text{t}}}$ denotes the angular-domain unitary matrix for the dual polarized directions, and $\bm{B} \in \mathbb{C}^{N_{\text{t}}/2 \times N_{\text{t}}/2}$ is the discrete Fourier transform (DFT) matrix. $\bm{F}_\text{D} \in \mathbb{C}^{M \times M}$ is the delay-domain DFT matrix. By exploiting the angular-delay-domain partial reciprocity between UL and DL channels, the BS selects $N_\text{p} \textless N_\text{t} M$ strongest ports in $\widetilde{\bm{H}}_{k, \text{UL}}$ to design DL pilots, where each port represents an angular-delay location. Besides, the corresponding indices of the selected ports $\{r_k(i), c_k(i)\}$, $i \in \{1, 2, ..., N_\text{p}\}$ are stored for the following reconstruction, wherein $r_k(i)$ and $c_k(i)$ denote the angular index and the delay index of the $i$-th selected port in $\widetilde{\bm{H}}_{k, \text{UL}}$. 

\textbf{Quantization}: At the $k$-th UE, after estimating the coefficient vector of the selected ports $\bm{w}_{k, \text{DL}} \in \mathbb{C}^{N_\text{p} \times 1}$, a quantization operation $Q(\cdot)$ is adopted to reduce the feedback overhead. Specifically, the $Q_\text{n}$-bit logarithmic quantization is firstly performed for normalizing two polarization directions. Then, for each port coefficient, the $Q_\text{a}$-bit logarithmic and $Q_\text{p}$-bit uniform quantizations are executed for the amplitude and phase, respectively. Finally, the quantized coefficient vector $Q(\bm{w}_{k, \text{DL}})$ is fed back to the BS. 

\textbf{Reconstruction}: At BS side, the recovered DL CSI matrix of the $k$-th UE in the angular-delay domain $\widehat{\widetilde{\bm{H}}}_{k, \text{DL}} \in \mathbb{C}^{N_\text{t} \times M}$ is initialized as an all-zero matrix. Then, the quantized coefficients in $Q(\bm{w}_{k, \text{DL}})$ are filled into their corresponding ports, which can be expressed as
\begin{equation}\label{reconstruction} 
\widehat{\widetilde{\bm{H}}}_{k, \text{DL}}[r_k(i), c_k(i)] = Q(\bm{w}_{k, \text{DL}})[i], i \in \{1, 2, ..., N_\text{p}\},
\end{equation}
where $\widehat{\widetilde{\bm{H}}}_{k, \text{DL}}[r_k(i), c_k(i)]$ denotes the coefficient in the $r_k(i)$-th row and the $c_k(i)$-th column of $\widehat{\widetilde{\bm{H}}}_{k, \text{DL}}$, and $Q(\bm{w}_{k, \text{DL}})[i]$ is the $i$-th coefficient in $Q(\bm{w}_{k, \text{DL}})$. Finally, the reconstructed DL CSI matrix in the spatial-frequency domain $\widehat{\bm{H}}_{k, \text{DL}} \in \mathbb{C}^{N_\text{t} \times M}$ can be acquired by the inverse transformation of (\ref{Eq:UL_DFT}).

\vspace*{-2mm}

\section{Deep Learning Assisted CSI Feedback \\ with TypeII Codebook}\label{Sec:Proposed} 

\subsection{Motivation}\label{S3.1} 

In spite of reducing the overhead of DL pilots and UL feedback, the latest TypeII codebook does not adequately utilize the correlations among the port coefficients, which could limit its performance. Specifically, the quantization step considers each port coefficient separately, but the correlations among the port coefficients are neglected. Moreover, only the coefficients of the selected ports are recovered in the reconstruction step. However, since the coefficients of the unselected ports that can be recovered based on the correlations are set to zero simply, the quality of CSI reconstruction is degraded.
 
To address these issues, considering that deep learning enjoys strong fitting capability to extract the correlations among the port coefficients, it is desirable to integrate the TypeII codebook with deep learning. Consequently, a tailored AE structure named TypeII-CsiNet is proposed for improving the performance of the TypeII codebook.

\vspace*{-2mm}

\subsection{Proposed TypeII-CsiNet}\label{S3.2} 

\begin{figure}[tp!]
\vspace*{-1mm}
\begin{center}
\includegraphics[width=.49\textwidth]{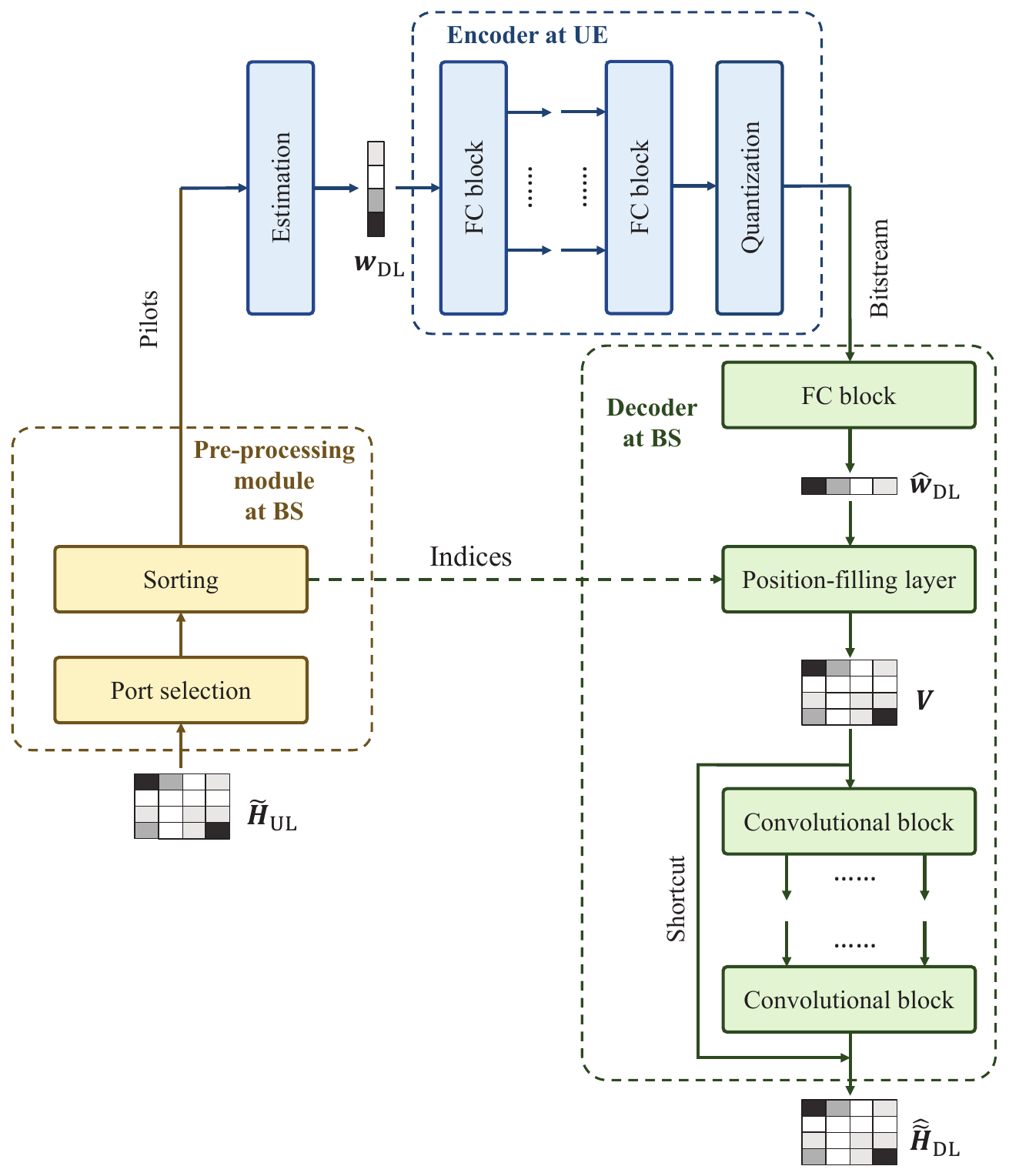}
\end{center}
\vspace{-4mm}
\caption{Structure of proposed TypeII-CsiNet.}
\label{Fig:TypeII-CsiNet} 
\vspace{-5mm}
\end{figure}

The proposed TypeII-CsiNet can be divided into three parts, the pre-processing module, encoder, and decoder, as shown in Fig.~\ref{Fig:TypeII-CsiNet}. Note that since the parameters of the encoders among different UEs are usually the same\cite{16}, the UE subscript $k$ is omitted in this subsection. 

\subsubsection{Pre-processing module}

At BS side, after carrying out the port selection step, the selected ports need to be further sorted to reserve the correlations in the coefficient vector $\bm{w}_\text{DL}$. Specifically, the port coefficients have the correlations in the angular-delay domain \cite{17,18}, which implies that they have the potential to be efficiently compressed by the AE \cite{5,10}. However, in the latest TypeII codebook, since various selected ports are shifted to the specific angular-delay locations for DL coefficient measurement \cite{2}, these selected ports in the CSI matrix are unknown to the UE, which destroys the stability of the correlations in $\bm{w}_\text{DL}$. This characteristic could hamper the performance of the AE, since deep learning is not good at extracting the features from non-sparse data \cite{13,14,15}. Consequently, to efficiently utilize the correlations, the selected ports need to be shifted to the stable positions in $\bm{w}_\text{DL}$, and we will elaborate the sorting methods together with the specific reasons in the following.


\textbf{Sorting by angle/delay indices}: The angular-delay-domain CSI matrix often exhibits the block sparsity, where the port coefficients from the same scatterer are clustered in some sub-blocks and have strong correlations \cite{5,19}. To reserve these correlations, the BS can sort the selected ports by the corresponding indices in the angular-delay-domain CSI matrix, so that the coefficients within the same sub-block can be clustered in $\bm{w}_\text{DL}$. Specifically, the BS can firstly sort the selected ports by their angular indices. Then, the ports within the same angular index are further sorted by the other index (their delay indices). For convenience, the above sorting method is termed as the sorting by angular-delay indices, while the sorting by delay-angular indices can be implemented in a similar way.

\textbf{Sorting by amplitudes}: In CSI feedback, the ports having larger amplitudes dominate the essential information of $\bm{w}_\text{DL}$, which could play a more crucial role than the weaker ports for the AE. Therefore, the stability of the correlations in amplitudes is important for the encoder to effectively compress $\bm{w}_\text{DL}$. Furthermore, considering that the coefficients of the selected ports are quite various for diverse UEs, sorting by amplitudes can strictly reserve the corresponding correlations of amplitude orders in $\bm{w}_\text{DL}$ for better compression. However, the amplitudes in $\bm{w}_\text{DL}$ are unknown to the BS. To cope with this issue, by exploiting the angular-delay-domain partial reciprocity between UL and DL channels, we propose to sort the selected ports by their corresponding amplitudes in the UL angular-delay-domain CSI matrix $\widetilde{\bm{H}}_\text{UL}$. 


\subsubsection{Encoder} 

At UE side, the estimated coefficient vector $\bm{w}_\text{DL}$ is firstly split into its real and imaginary parts, and concatenated to $\big[\Re(\bm{w}^\text{T}_\text{DL}), \Im(\bm{w}^\text{T}_\text{DL}\big)]^\text{T} \in \mathbb{C}^{2N_\text{p} \times 1}$ for feature extraction. Then, multiple fully-connected (FC) blocks are deployed to learn the correlations for compression. Finally, the $Q_\text{u}$-bit uniform quantization is performed for each output unit of the last FC block, which forms the feedback bitstream with the length $B$. It is noted that the quantization is non-differentiable, hence stacking sigmoid functions is introduced for approximating the quantization gradient \cite{7}.


\subsubsection{Decoder} 

At BS side, a FC block is firstly used for the dimension recovery from the feedback bitstream to the coarse reconstruction of the coefficient vector $\widehat{\bm{w}}_\text{DL} \in \mathbb{C}^{N_\text{p} \times 1}$. After that, the coefficients in $\widehat{\bm{w}}_\text{DL}$ are filled into their corresponding ports in the initialized all-zero CSI matrix $\bm{V} \in \mathbb{C}^{N_\text{t} \times M} $ by the position-filling layer. In this way, the angular-delay-domain structure of the recovered CSI matrix is adequately leveraged, which brings two advantages for the subsequent reconstruction. Firstly, it is beneficial to sufficiently leverage the angular-delay-domain correlations to accurately refine the coefficients of the selected ports. Secondly, based on the correlations, the coefficients of the unselected ports can be recovered according to the selected ports. Finally, multiple convolutional blocks with the shortcut are adopted for the reconstruction of the angular-delay-domain CSI matrix $\widehat{\widetilde{\bm{H}}}_{\text{DL}}$.

\vspace*{-3mm}

\subsection{Training of Proposed TypeII-CsiNet}\label{S3.3} 

In the training of our proposed TypeII-CsiNet, the two-stage loss function is designed to sufficiently boost the sum rate performance in MU-MIMO scenarios. Before introducing it, we briefly present two existing loss functions. Firstly, the MSE loss function $L_\text{MSE}$ is one of the most common loss functions for CSI feedback in single-UE scenarios \cite{3,4,5}, which can be written as
\begin{equation}\label{Eq:MSE} 
L_\text{MSE} = \frac{1}{K} \sum_{k=1}^{K} \big\|\widetilde{\bm{H}}_{k, \text{DL}}-\widehat{\widetilde{\bm{H}}}_{k, \text{DL}}\big\|^2,
\end{equation}
where $\widetilde{\bm{H}}_{k, \text{DL}} \in \mathbb{C}^{N_\text{t} \times M}$ is the DL angular-delay-domain CSI matrix containing perfect coefficients of both selected and unselected ports for the $k$-th UE. Secondly, the negative average rate loss function $L_\text{NAR}$ \cite{16,20,21} tailored for the unsupervised optimization in MU-MIMO scenarios can be expressed as
\begin{equation}\label{Eq:NAR} 
L_\text{NAR}= -\frac{1}{K}R_\text{SUM}.
\end{equation}
In this paper, we consider CSI feedback in MU-MIMO scenarios, hence $L_\text{NAR}$ is the intuitive loss function for improving the sum rate performance. However, since $L_\text{NAR}$ only provides a low-dimensional scalar guide during model optimizations, $L_\text{NAR}$ guided training suffers from many local optimums. On the contrary, $L_\text{MSE}$ focuses on the reconstruction of the CSI matrix for each UE separately, and $L_\text{MSE}$ guided training only has a global optimum. Nevertheless, the suppression of MU interference is ignored in $L_\text{MSE}$. Thus, using $L_\text{NAR}$ or $L_\text{MSE}$ in isolation could degrade the sum rate performance. 

Fortunately, we observe that $L_\text{NAR}$ and $L_\text{MSE}$ are complementary from the above discussion. Specifically, $L_\text{MSE}$ can assist $L_\text{NAR}$ in avoiding the trapping in local optimums, while $L_\text{NAR}$ can contribute to the capability of MU interference suppression for $L_\text{MSE}$. Thus, the mixed loss function $L_\text{MIX}$ is proposed to integrate the advantages of $L_\text{MSE}$ and $L_\text{NAR}$ as below
\begin{equation}\label{Eq:Mixed} 
L_\text{MIX} = L_\text{MSE} + \mu L_\text{NAR},
\end{equation}
where $\mu$ is the non-negative hyperparameter. Based on the proposed mixed loss function, we further propose a two-stage loss function to more adequately avoid the trapping in local optimums. Concretely, $L_\text{MSE}$ is used for approaching the perfect CSI label in the first stage. Then, in the second stage, $L_\text{MIX}$ is used for further improving the sum rate performance.

\vspace*{-1mm}

\section{Simulation Results}\label{Sec:Sim} 

\addtolength{\topmargin}{0.03in}

\subsection{Simulation System Setup}\label{S4.1} 

\begin{table}[tp!] 
\vspace*{-5mm}
    \begin{center}
        \caption{System parameters.}
        \vspace*{-2mm}
        \setlength{\tabcolsep}{2mm}
        \begin{tabular}{cc}
            \toprule[0.8pt]
            Parameters                                             & Values                    \\
            \toprule[0.8pt]
 		  Center frequency for UL/DL                                         & $3.4/3.5$\,GHz              \\
            Cell radius                                                     & $250$\,m                \\
            DL transmit power $P$                                           & $35$\,dBm            \\
            Noise factor                                                    & $5$\,dB                \\
            Number of UEs $K$                                          & $5$                     \\
            Numbers of BS horizontal/vertical antennas $N_\text{h}/N_\text{v}$            & $4/4$                     \\
            Number of BS antennas $N_\text{t}$                              & $32$                    \\
		 Subcarrier spacing                                              & $15$\,kHz              \\
            RB spacing                                                      & $180$\,kHz              \\
            Number of subbands $M$                                            & $16$             \\
            \toprule[0.8pt]
            \vspace{-9mm}
        \end{tabular}
        \label{tab1}
    \end{center}
\end{table}

A single-cell FDD MU-MIMO system is considered in the simulations, where the channel model of 3GPP TR 38.901 for urban macro-cell scenarios is used to generate the data set \cite{22}, and the corresponding parameters are listed in Table~\ref{tab1}. For simplicity, the zero-forcing precoding is employed for simultaneously serving all UEs \cite{23}.

\begin{table}[tp!]
    \begin{center}
        \caption{Proposed deep learning structures.}
        \vspace*{-2mm}
        \setlength{\tabcolsep}{2mm}
        \begin{tabular}{ccc}
            \toprule[0.8pt]
            Modules         & Blocks                     & Parameters  \\
            \toprule[0.8pt]
            \multirow{2}{*}{Encoder}  
                & FCB 1 & $f_\text{i}=2N_\text{p},f_\text{o}=1024$, BN, LReLU(0.3)\\
                & FCB 2 & $f_\text{i}=1024,f_\text{o}=\lfloor B/2 \rfloor$, BN, Tanh\\
            \midrule[0.8pt]
            \multirow{4}{*}{Decoder}  
                & FCB & $f_\text{i}= \lfloor B/2 \rfloor,f_\text{o} = 2N_\text{p}$, dropout(0.02)\\
                & CB 1 & $f_\text{i}=2,f_\text{o}=128$, BN, LReLU(0.1)\\
                & CB 2-9 & $f_\text{i}=128,f_\text{o}=128$, BN, LReLU(0.1)\\
			& CB 10 & $f_\text{i}=128,f_\text{o}=2$\\
            \toprule[0.8pt]
            \vspace{-12mm}
       \end{tabular}
       \label{tab2}
    \end{center}
\end{table}

For the proposed TypeII-CsiNet, the network parameters are specified in Table~\ref{tab2}. In particular, FCB and CB denote the FC and convolutional blocks, respectively, wherein $f_\text{i}$ and $f_\text{o}$ represent the numbers of input and output feature channels, BN denotes the batch normalization layer, and LReLU and Tanh are the leaky reflected linear unit and hyperbolic tangent activation functions. Besides, the two-dimensional kernel size, stride, and padding size in the CB are $(3,3)$, $(1,1)$, and $(1,1)$, respectively. For the uniform quantization in the encoder, the length of bits for each output unit $Q_\text{u}$ is set to $2$. Furthermore, the training data set and test data set consist of $107520$ and $5120$ samples, respectively. The batch size is set to $32$. The Adam optimizer with the initial learning rate $lr = 3 \times 10^{-2}$ is adopted. The whole training process contains $200$ epochs. For the two-stage loss function, the first stage contains the first $80$ epochs and the remaining epochs are the second stage. In addition, $\mu$ is set to $1 \times 10^{-3}$ for the mixed loss function.

\begin{figure}[tp!]
\begin{center}
\includegraphics[width=.47\textwidth]{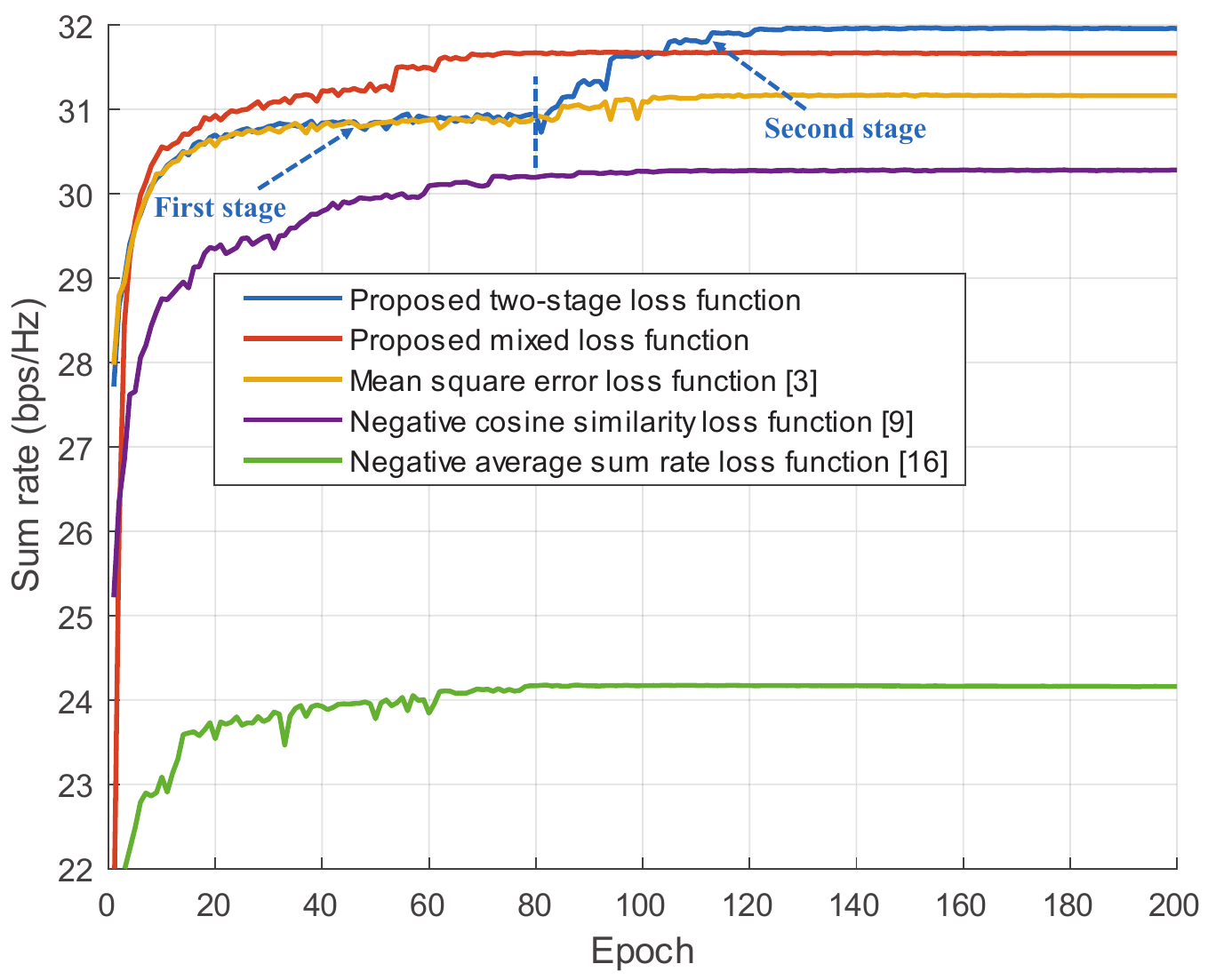}
\end{center}
\vspace{-4mm}
\caption{Convergence performance of TypeII-CsiNet over different loss functions with $N_\text{R}=4$, $N_\text{p}=32$, and $B=165$.}
\label{Fig:loss_functions} 
\vspace{-4mm}
\end{figure}

\vspace*{-2mm}
\subsection{Performance Analysis}\label{S4.2} 

Firstly, Fig.~\ref{Fig:loss_functions} shows the convergence performance of our proposed TypeII-CsiNet for different loss functions including MSE \cite{3}, negative cosine similarity \cite{9}, negative average rate \cite{16}, and the proposed loss functions, where the sorting by amplitudes is adopted for a fair comparison. The number of RBs within one subband $N_\text{R}$, the number of selected ports $N_\text{p}$, and the length of feedback bits $B$ are set to $4$, $32$, and $165$, respectively. It can be seen that the proposed mixed loss function achieves better sum rate performance than the existing loss functions, which verifies that combining the MSE and negative average rate loss functions is beneficial to optimizing TypeII-CsiNet. Furthermore, it can be observed that the two-stage loss function achieves the best sum rate performance, which demonstrates that the MSE loss function applied in the first stage is beneficial to more sufficiently avoiding the trapping in local optimums. Therefore, we use the two-stage loss function to train our proposed TypeII-CsiNet for the following simulations.

\begin{figure}[tp!]
\begin{center}
\includegraphics[width=.47\textwidth]{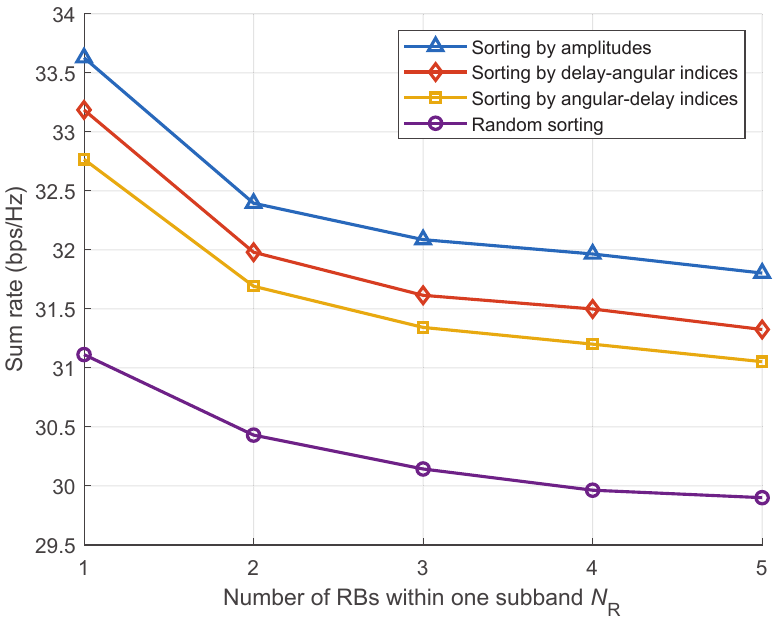}
\end{center}
\vspace{-4mm}
\caption{Sum rate performance comparison of TypeII-CsiNet for different sorting methods as function of $N_\text{R}$ with $N_\text{p}=32$ and $B=165$.}
\label{Fig:sort} 
\vspace{-4mm}
\end{figure}

Next, the sum rate performance of different sorting methods with respect to the number of RBs within one subband $N_\text{R}$ is compared in Fig.~\ref{Fig:sort}, wherein the random sorting is depicted to validate the necessity of sorting. The number of selected ports $N_\text{p}$ and the length of feedback bits $B$ are set to $32$ and $165$, respectively. Obviously, as $N_\text{R}$ increases, we can see that the sum rate performance of all sorting methods is degraded accordingly. This is because the frequency correlations of the port coefficients become weaker due to the larger bandwidth. Moreover, it can be observed that the sum rate performance of the three sorting methods significantly surpasses the random sorting, which shows that sorting the selected ports can effectively facilitate the feedback with the AE. Besides, the sorting by amplitudes achieves the best sum rate performance in all $N_\text{R}$, which indicates that the correlations of amplitude orders are more beneficial for the compression by the AE.

\vspace*{-2mm}

\subsection{Performance Comparison}\label{S4.3} 

In this subsection, we further compare our proposed TypeII-CsiNet with the TypeII codebook and existing deep learning benchmarks. In the TypeII codebook \cite{2}, for compressing the port coefficients, the quantization bits for normalizing two polarization directions $Q_\text{n}$ are set to $5$, while the quantization bits for the amplitude and phase of each port coefficient $\{(Q_\text{a}, Q_\text{p})\}$ are set from $\{(1, 2), (2, 2), (2, 3), (3, 3), (3, 4)\}$. For deep learning benchmarks, we illustrate the TypeII-CsiNet without the position-filling layer, and EVCsiNet \cite{9} is adopted to represent the existing AE. In these two benchmarks, the encoder compresses the coefficient vector of the selected ports, and the decoder directly reconstructs the corresponding coefficient vector instead of the whole CSI matrix. Besides, for fair comparisons, our proposed TypeII-CsiNet and these deep learning benchmarks all adopt the sorting by amplitudes.

\begin{figure}[tp!]
\begin{center}
\includegraphics[width=.47\textwidth]{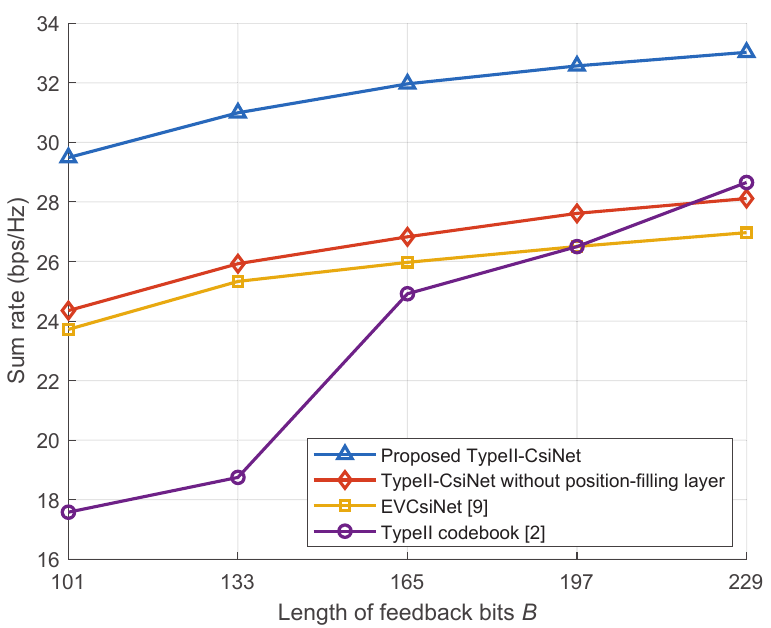}
\end{center}
\vspace{-4mm}
\caption{Sum rate performance comparison of different methods as function of $B$ with $N_\text{R}=4$ and $N_\text{p}=32$.}
\label{Fig:bits} 
\vspace{-4mm}
\end{figure}

In Fig.~\ref{Fig:bits}, we compare the sum rate performance of different methods as the function of the length of feedback bits $B$, which can be calculated by $B=Q_\text{n}+N_\text{p}\times(Q_\text{a}+Q_\text{p})$. The number of RBs within one subband $N_\text{R}$ and the number of selected ports $N_\text{p}$ are set to $4$ and $32$, respectively. Firstly, it can be seen that the TypeII-CsiNet without the position-filling layer outperforms EVCsiNet in all $B$, which verifies the effectiveness of our proposed two-stage loss function for CSI feedback in MU-MIMO scenarios. Then, a substantial improvement is brought further by the position-filling layer in the proposed TypeII-CsiNet, which demonstrates that the position-filling layer leveraging the angular-delay-domain structure of the recovered CSI matrix can significantly improve the reconstruction accuracy. Besides, for the TypeII codebook, we can see that the curve rises sharply from $133$ bits to $165$ bits and from $197$ bits to $229$ bits, which indicates that its performance is more sensitive to the quantization bits for the phase $Q_\text{p}$. Finally, we can observe that the proposed TypeII-CsiNet achieves more significant enhancements in the cases with lower feedback overhead compared to the TypeII codebook. It demonstrates that the proposed TypeII-CsiNet can sufficiently exploit the correlations among the port coefficients to enhance the compression performance.

\begin{figure}[tp!]
\begin{center}
\includegraphics[width=.47\textwidth]{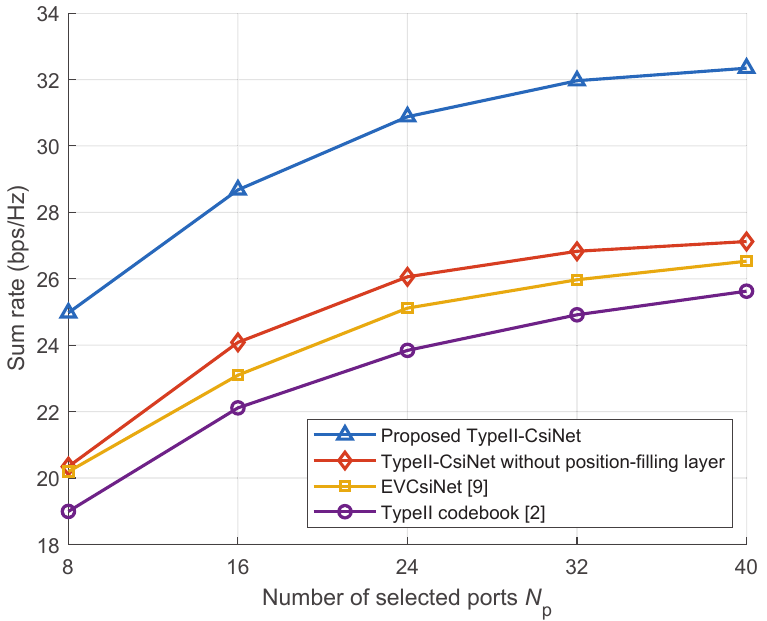}
\end{center}
\vspace{-4mm}
\caption{Sum rate performance comparison of different methods as function of $N_\text{p}$ with $N_\text{R}=4$ and $B=165$.}
\label{Fig:Port} 
\vspace{-4mm}
\end{figure}

Finally, the sum rate performance as the function of the number of the selected ports $N_\text{p}$ is investigated for different methods in Fig.~\ref{Fig:Port}. The number of RBs within one subband $N_\text{R}$ and the length of feedback bits $B$ are set to $4$ and $165$, respectively. Firstly, it is clear that the sum rate performance is improved as $N_\text{p}$ increases due to reserving more adequate information of DL CSI. In addition, the performance of the proposed TypeII-CsiNet is significantly superior to other methods in all $N_\text{p}$, which demonstrates that our proposed TypeII-CsiNet can effectively utilize the correlations among the port coefficients to accurately reconstruct the whole CSI matrix for enhancing the TypeII codebook.

\section{Conclusions}\label{Sec:CONCLUSION}

In this paper, we propose a tailored AE structure named TypeII-CsiNet to effectively integrate the TypeII codebook with deep learning, wherein three novel designs are developed. Firstly, to address the instability of the correlations among the feedback coefficients, the block sparsity and amplitude orders are exploited to sort the selected ports. Secondly, a position-filling layer is developed in the decoder to fill the feedback coefficients into their ports in the recovered CSI matrix, so that its angular-delay-domain structure is fully leveraged to enhance the reconstruction accuracy. Thirdly, we analyze the MSE and negative average rate loss functions, and a two-stage loss function is proposed to integrate their advantages for sufficiently boosting the sum rate performance. Simulation results have verified that our proposed TypeII-CsiNet outperforms the TypeII codebook and existing deep learning benchmarks in terms of the sum rate performance. 

\section*{Acknowledgment}

This work was supported in part by the National Natural Science Foundation of China under Grant U22B2057 and in part by Huawei Research Fund. (\emph{Corresponding author: Zhaocheng Wang})


{

}

\end{document}